# X-ray Photoelectron Spectroscopy Studies of $MgB_2$ for Valence State of Mg

**A.Talapatra[1], S.K.Bandyopadhyay[1*], Pintu Sen[1] and P.Barat[1]**
**S.Mukherjee[2] and M.Mukherjee[2]**

**1. Variable Energy Cyclotron Centre,**
**1/AF, Bidhan Nagar, Kolkata-700 064, India.**
**2. Saha Institute of Nuclear Physics**
**1/AF, Bidhan Nagar, Kolkata-700 064, India.**

**Abstract:**

Core level X-ray Photoelectron Spectroscopy (XPS) studies have been carried out on polycrystalline $MgB_2$ pellets over the whole binding energy range with a view to having an idea of the charge state of Magnesium (Mg). We observe 3 distinct peaks in Mg 2p spectra at 49.3 eV (trace), 51.3 eV (major) and 54.0 eV (trace), corresponding to metallic Mg, $MgB_2$ and $MgCO_3$ or, divalent Mg species respectively. Similar trend has been noticed in Mg 2s spectra. The binding energy of Mg in $MgB_2$ is lower than that corresponding to Mg(2+), indicative of the fact that the charge state of Mg in $MgB_2$ is less than (2+). Lowering of the formal charge of Mg promotes the $\sigma \rightarrow \pi$ electron transfer in Boron (B) giving rise to holes on the top of the $\sigma$-band which are involved in coupling with B $E_{2g}$ phonons for superconductivity. Through this charge transfer, Mg plays a positive role in hole superconductivity. B 1s spectra consist of 3 peaks corresponding to $MgB_2$, boron and $B_2O_3$. There is also evidence of MgO due to surface oxidation as seen from O 1s spectra.

**PACS No. 74.70.Ad**

**Keywords: $MgB_2$, XPS, Binding Energy.**

• **Corresponding author. E-mail address: skband@veccal.ernet.in**

**Introduction:**

Since the inception of Magnesium Diboride ($MgB_2$) superconductor [1], it has attracted a lot of interest because it is a simple binary intermetallic system with fairly high $T_c$ of ~39K. The relatively high $T_c$ and high conductivity place it among potential attractive materials for application in devices. $MgB_2$ is within the clean limit with large coherence length (~5 nm) as compared to HTSC cuprates. The flow of supercurrents in $MgB_2$ is not significantly hindered by grain boundaries [2] which is dominant in cuprates. The observation of significant boron isotope effect on superconducting transition temperature $T_c$ [3] and heat capacity measurement [4] indicate that $MgB_2$ is a phonon mediated s-wave [5] superconductor.

Experimental results suggest that it is a multiband superconductor [6]. $MgB_2$ is built up on hexagonal structure where magnesium (Mg) layers are sandwiched between two boron (B) layers. There is a strong $sp^2$ hybrid $\sigma$ bonding within in-plane boron atoms that gives rise to 2D $\sigma$ bands. The out-of-plane B atoms are coupled by $p_z$ orbitals through Mg atoms giving rise to 3D metallic $\pi$ bands. It is important to have an idea about the charge state of Mg for full understanding of the superconductivity in this system. Binding energy (B.E.) of core level and valence electrons of Mg can highlight the valence state i.e. whether it is (2+) or different. XPS is a tool to probe the binding energies of core electrons.

There have been some XPS studies with respect to binding energy aspects in this system [7-9] but the values differ significantly. In this paper, we are reporting some XPS studies of the polycrystalline $MgB_2$. We have analysed the spectra of Mg 2p and B 1s with a view to having an idea of the charge state of Mg. We have also taken the spectra of O 1s to have the idea of surface oxygen and traces of MgO.

**Experimental:**

$MgB_2$ pellets have been synthesized starting from Magnesium (> 97% pure) and Boron (99.9% pure) powder with 2 atomic percent Mg in excess to stoichiometry and heating them in an inert atmosphere of helium in a sealed tube at 800-900$^0$C [10]. Excess Mg was employed to compensate for the loss of volatile Magnesium. The mixture was pelletised by applying uniaxial pressure of 7.5 tons/cm$^2$. The pellets were wrapped in Tantalum foil, put in quartz tube and evacuated to $10^{-2}$ torr. They were purged with 99.9% pure helium quite a number of times to expel the oxygen in air and sealed in helium environment (at ~ 800 torr). The inert gas environment was needed to avoid the formation of MgO phase. The samples were heated in quartz tube at 800$^0$C for two hours and then at 900$^0$C for one hour and quenched at 625$^0$C to arrest the phase of $MgB_2$. Formation of $MgB_2$ is exothermic and hence the employment of tantalum foil as heat sink was essential. The samples thus obtained were characterized by X-ray diffraction (XRD) taken with Philips PW1710 diffractometer with Cu $K_\alpha$ of wavelength 1.54 Å. Resistivities of the samples as a function of temperature were measured in close cycle helium refrigerator (made of Cryoindustries of America) using four probe technique with HP 34220A Nanovoltmeter with resolution of 0.1 nanovolt and Keithley 224 programmable current source. 1mA current was employed.

XPS measurements were carried out on one sample with Omicron NanoTechnology MXPS instrument using monochromatised Al-K$\alpha$ radiation with energy of 1486.7eV and resolution of 0.6 eV. The sample was first sonicated with ethanol and its surface was cleaned by repeated sputtering with inert ultrahigh pure Ar-gas in-situ. The base pressure of the chamber was $8\times10^{-10}$ torr and the XPS data were collected at $2\times10^{-9}$ torr. Energy calibration was done with Ag-3$d_{5/2}$ line of B.E. 368.2 eV. The survey scan was taken for the entire binding energy. The peak positions in individual spectra were determined using the linear background and least square fitting of the data to Gaussian line shape.

**Results and Discussions:**

Fig. 1 shows the XRD pattern of the polycrystalline $MgB_2$ pellet prepared by us. All reflection planes characteristic of $MgB_2$ have been obtained. The lattice parameters were obtained by indexing the reflection planes with Rietveld analysis using LS1 programme. The values obtained, a=3.09 Å and c= 3.49 Å, are in good agreement with the literature values [1]. Fig. 2 depicts the plot of resistivity versus temperature for $MgB_2$. Tc (R=0) was observed to be 38.7K. A sharp transition was noticed with the width ($\Delta$Tc) of 0.6K indicative of strong intergranular coupling. Residual resistivity ratio defined as R(300K)/R(40K) was around 3.

The XPS survey scan is displayed in fig. 3. We observe distinct peaks from Mg, B and O. A small peak of C 1s (coming from the surface contamination) was observed even after repeated bombardment with $Ar^+$ of 1 KeV. The respective spectra are being discussed later. The Mg KLL Auger peaks in the region of 300-380 eV correspond to Auger signals of kinetic energies in the region 1106-1186 eV. Aswal et al [7] observed Auger signals of kinetic energies 1186 and 1180 eV. The Oxygen KVV Auger peaks at 980.0 eV correspond to Auger kinetic energy of 506.7 eV.

As prepared with slight excess of Mg to prevent decomposition, $MgB_2$ pellet is highly vulnerable to surface oxidation as observed in O 1s spectra. The surface contained oxygen (as oxide) as also seen by Garg et al. [9] though it decreased considerably after argon sputter cleaning of the surface. In fig. 4, we have plotted the peak intensity ratio of the Mg 1s to O 1s peak as a function of the sputtering time. Increase of this ratio is a measure of the decrease of the oxygen content due to sputtering. We can clearly see that the oxygen concentration almost saturates after sputtering for 120 minutes with $Ar^+$ of 1 KeV. X-Ray Absorption Spectroscopy (XAS) and X-ray Emission Spectroscopy (XES) studies by Mc Guiness et al [11] on polycrystalline $MgB_2$ also unmistakably reveal structures arising from oxides.

Fig. 5 depicts the fitted Mg 2p core level spectra containing the observed one with raw data, fitted and deconvoluted to resolve the peaks. Mg 2p level being close to valence band is expected to reveal the charge and valence state of Mg more representatively. The difference between Mg $2p_{1/2}$ and $2p_{3/2}$ is not resolvable. We are able to fit three distinct Gaussian components in case of the sputtered sample. The peak at 49.3 eV corresponds to the metallic Mg which was also observed by other groups [8,9]. Vasquez et al. [8] assigned it to be $MgB_2$, but it has been established later [9] to correspond to metallic Mg. The major peak at 51.3 eV corresponds to $MgB_2$ [9]. Normally one would expect a core level shift towards higher binding energies (i.e. positive shift) in losing valence electron, because less electrons screen the coulomb potential and therefore, the core electrons are bound stronger. Following the similar argument, anions would have lower binding energy as compared to normal atoms. Thus, we can infer that the minor peak at higher energy of ~54.0 eV corresponds to $MgCO_3$ and/or dioxygen species Magnesium peroxide with Mg as (2+) [9]. Hence, the binding energy of Mg in $MgB_2$ is less than that corresponding to Mg(2+), indicative of the fact that the charge state of Mg in $MgB_2$ is less than (2+). Mg was earlier believed to be as (2+) in this compound. Electrons are donated to boron. In this context, it would be ideal to have a look at the structure and bonding in $MgB_2$. $MgB_2$ is structurally similar to $AlB_2$. B atoms reside in graphite like honeycomb layers with Mg atoms sandwiched between B layers. This structure may therefore, be regarded as that of the intercalated graphite [12] with carbon replaced by B. Furthermore, $MgB_2$ is formally isoelectronic with graphite. Therefore, chemical binding and electronic properties of $MgB_2$ are expected to bear similarity to those of graphite and intercalated compounds, some of which show superconductivity. However, in spite of the structural similarity to the intercalated graphite, $MgB_2$ is qualitatively different from the former. The unique feature of $MgB_2$ is the incomplete filling of two $\sigma$-bands arising from $sp^2$ hybridisation of 2s and two 2p orbitals of B which is being explained later. In isoelectronic and isostructural graphite and $AlB_2$, $\sigma$-states are completely filled with strong covalent bonds [13]. In $MgB_2$, the holes are localised within 2 dimensional B sheets, in contrast with mostly 3 dimensional electrons and holes in the $\pi$ bands, which are delocalised over the whole crystal. These 2D covalent and 3D metallic type states contribute to the density of states (DOS) at the Fermi level, while the unfilled covalent bands experience strong interaction with longitudinal vibrations in the B layer [13,14].

$MgB_2$ is in another way very unique in contrast to $MgF_2$, $MgCO_3$ or, Mg-peroxide which are almost ionic in nature with Mg as (2+). Fluorine or, oxygen being more electronegative than B causes almost full charge transfer from Mg. On the contrary, in $MgB_2$, there is likelihood of some back charge transfer from B to Mg(2+) leading to a lowering of the formal charge of Mg and hence the partial covalency in Mg-B bond. Mg is not fully ionic. On the same ground, even in $AlB_2$, the complete charge transfer of three electrons to B does not occur. However, with partial charge transfer also, in $AlB_2$, the σ-bands are completely filled. The partial charge transfer from B to Mg is analogous to that from Oxygen to Cu in Cuprates [15], whereby holes reside in oxygen. Superconductivity in cuprates is driven by hole transfer through the planar pπ orbitals [16].

In $MgB_2$, the bonding situation is as follows. Mg stacked between two boron layers donates electrons to them. The electrons go to B π bands, as in-plane σ-bands are completely occupied. Due to strong attractive potential of $Mg^{2+}$, there is some electron transfer from B π to Mg (2+) causing a reduction of charge state of Mg from (2+). Due to the interaction between Mg and B, Mg 3s states are pushed up causing a lowering of B 2pπ band level, resulting in $\sigma \rightarrow \pi$ charge transfer that drives the hole doping (to the extent of ~0.17 holes) of the σ-bands [13,15,17]. These holes couple with B $E_{2g}$ mode phonons giving rise to superconductivity. This is favoured by the lowering of formal charge of Mg. This is a unique feature in $MgB_2$ which does not occur in the graphite, where σ-states are filled leading to the loss of superconductivity.

Fig. 6 depicts the fitted B 1s spectra. The peak at lower energy of 187.7 eV is the major one corresponding to $MgB_2$ showing the Boron as boride anion and is consistent with the B.E. values (187.2-188.5 eV) reported for transition metal diborides [18,19]. These indicate that the bonding between B-B is essentially covalent [19]. Formally, $MgB_2$ was assigned to be $Mg^{2+}(B_2)^{2-}$ putting an average charge of (-1) at B. But, if the charge in Mg is less than (2+), then the average charge on B will be less than (-1). However, the negative charge in B is delocalised over the entire lattice. Hence we cannot distinguish the effect of slight increase of the B.E. of B due to charge transfer to Mg, as there is a wide range of B.E.s in transition metal borides (187.2-188.5 eV). Charge density (CD) plots reveal that the bonding in B layer is essentially covalent [14]. The CD of B atom is strongly aspherical and the directional bonds with high CD are seen. However, a large amount of valence charge does not participate in any covalent bonding, but is rather distributed more or, less homogeneously over the whole crystal.

There is also significant amount of the component corresponding to the higher binding energy of 192.9 eV and a small peak at 190.3 eV corresponding to positively charged boron in $B_2O_3$ and metallic B respectively. McGuiness et al. [11] and Garg et al. [9] also reported the presence of $B_2O_3$ which we attribute to the surface oxidation, which was not completely removed even after long sputtering. Also, some unreacted B was remaining like the unreacted Mg observed in Mg 2p spectra.

Numbers obtained from Mg 2s spectra presented in Table-1 also reveal the patterns similar to Mg 2p spectra. In case of Mg 2s, there is a major central peak corresponding to $MgB_2$ at 90.1 eV added with two peaks at 88.0 eV and 93.0 eV, corresponding to metallic Mg and divalent Mg compounds respectively [20]. For Mg 1s spectra, the respective positions for $MgB_2$, metallic Mg and divalent Mg species are 1305.5 eV, 1304.0 eV and 1306.6 eV respectively [20]. Intensities of the peaks of Mg 1s spectra are much more compared to Mg 2s and Mg 2p peaks.

O 1s spectra in fig. 7 revealed two distinct peaks- the lower B.E. component at 532.4 eV corresponding to MgO at the surface and the higher one occurring at 535.1 eV corresponding to $B_2O_3$. Oxygen attached to B (more electronegative than Mg) shows higher B.E. of core electrons of Oxygen for the reasons mentioned earlier. This was also observed by Garg et al.[9] where the oxygen content decreased with argon sputtering but could not be eliminated for sintered polycrystalline granular samples.

**Conclusion:**

We have done XPS studies of polycrystalline $MgB_2$. The survey scan revealed peaks due to Mg 2p, Mg 2s, Mg 1s, B1s and considerable O 1s. Surface oxidation gave rise to peaks of MgO and $B_2O_3$. Analysis of Mg spectra (Mg 2p, Mg 2s and Mg 1s) revealed the presence of metallic Mg and divalent Mg (in form of $MgCO_3$ and/or Mg peroxide) in addition to $MgB_2$. The charge state of Mg appears to be between 0 and (2+) (less than 2+). Mg is not fully ionised here as was believed earlier. There is an electron transfer from B $2p_z$ orbitals to Mg (2+) causing a lowering of its formal charge. The interaction of Mg and B orbitals pushes up Mg 3s levels and lowers B 2p$\pi$ orbitals. This induces a charge transfer $\sigma\rightarrow\pi$ band in B giving rise to holes on the top of $\sigma$-band which couples with B $E_{2g}$ vibration mode phonons. This is the mechanism of superconductivity in $MgB_2$ which is favoured by the above mentioned charge transfer. Thus, in contrast to the isoelectronic graphite, and other transition metal borides (where $\sigma$-bands are completely filled and there is no existence of holes), in $MgB_2$, the initial charge transfer from Mg to B and then partial back charge transfer from B to Mg (analogous to that from O to Cu in Cuprates) play a positive role in superconductivity within the framework of electron phonon mechanism. B spectra reveal that the bonding of B is similar to that of transition metal borides, though it does not clearly resolve the effect of charge transfer. O 1s spectra show some presence of MgO due to surface oxidation which could not be fully eliminated even after repeated argon ion sputtering.

**Acknowledgements:**

Authors would like to thank Prof. B. Sinha, Director, VECC, for his constant encouragement. Authors also express thanks to A. Sarkar for Rietveld analysis. One of the authors, AT acknowledges CSIR, Govt. of India for financial assistance.

Figure Captions.

Fig. 1. X-ray Diffraction pattern of polycrystalline $MgB_2$. The pattern shows all the peaks due to $MgB_2$ with traces of MgO.

Fig. 2. Resistivity Vs.temperature plot of polycrystalline $MgB_2$ with a sharp transition indicating strong intergranular coupling.

Fig. 3. XPS survey scan of entire Binding energy.

Fig. 4. Intensity Ratio of Mg 1s to O 1s peak as a function of sputtering time.

Fig. 5. Mg 2p spectra. The dotted line denotes spectra with raw data, the solid line the fitted one and the dashed line the deconvoluted spectra displaying the major peak of $MgB_2$ at 51.3 eV and minor peaks at 49.3 eV and 54.0 eV representing metallic Mg and divalent species of Mg respectively.

Fig. 6. B 1s Spectra- the dotted line shows the spectra with raw data, the continuous line the fitted one and the dashed one gives the deconvoluted spectra. The lower energy peak at 187.6 eV represents $MgB_2$. the peak at 190.3 corresponds to some unreacted B and the one at 192.9 eV corresponds to $B_2O_3$ due to surface oxidation.

Fig. 7. O 1s spectra. The dotted, continuous and dashed lines representing the spectra with raw data, fitted and the deconvoluted spectra respectively. The peak at 532.4 eV corresponds to MgO and at 535.1 eV corresponds to $B_2O_3$.

Table-1. Parameters of Mg 2s spectra.

| Mg 2s | Peak energy (eV) | Species | % Area under the curve |
|---|---|---|---|
| | 88.0 | Metallic Mg | 5.5 |
| | 90.1 | $MgB_2$ | 83.5 |
| | 93.0 | $MgCO_3$ | 11.0 |

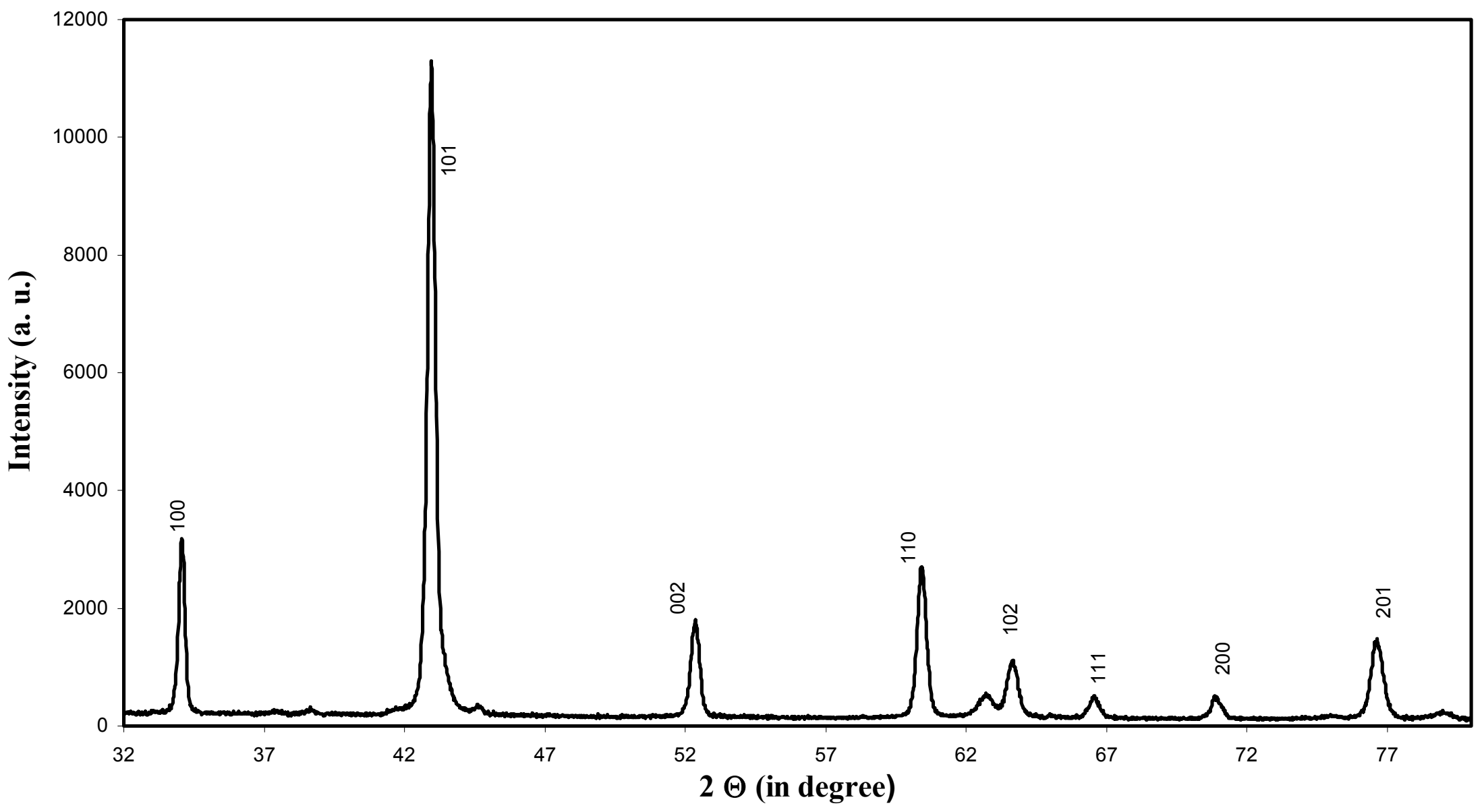

Fig.1

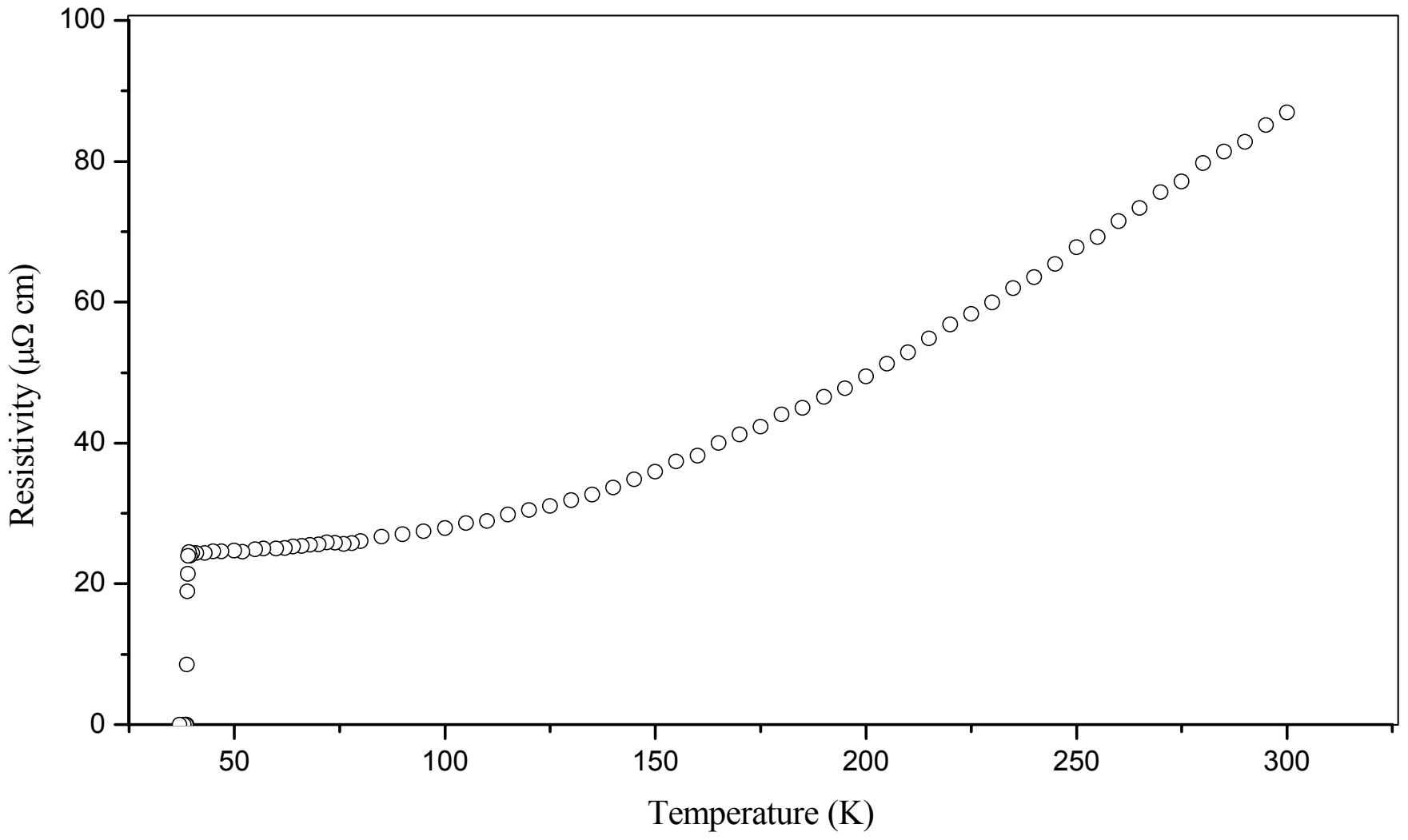

Fig. 2

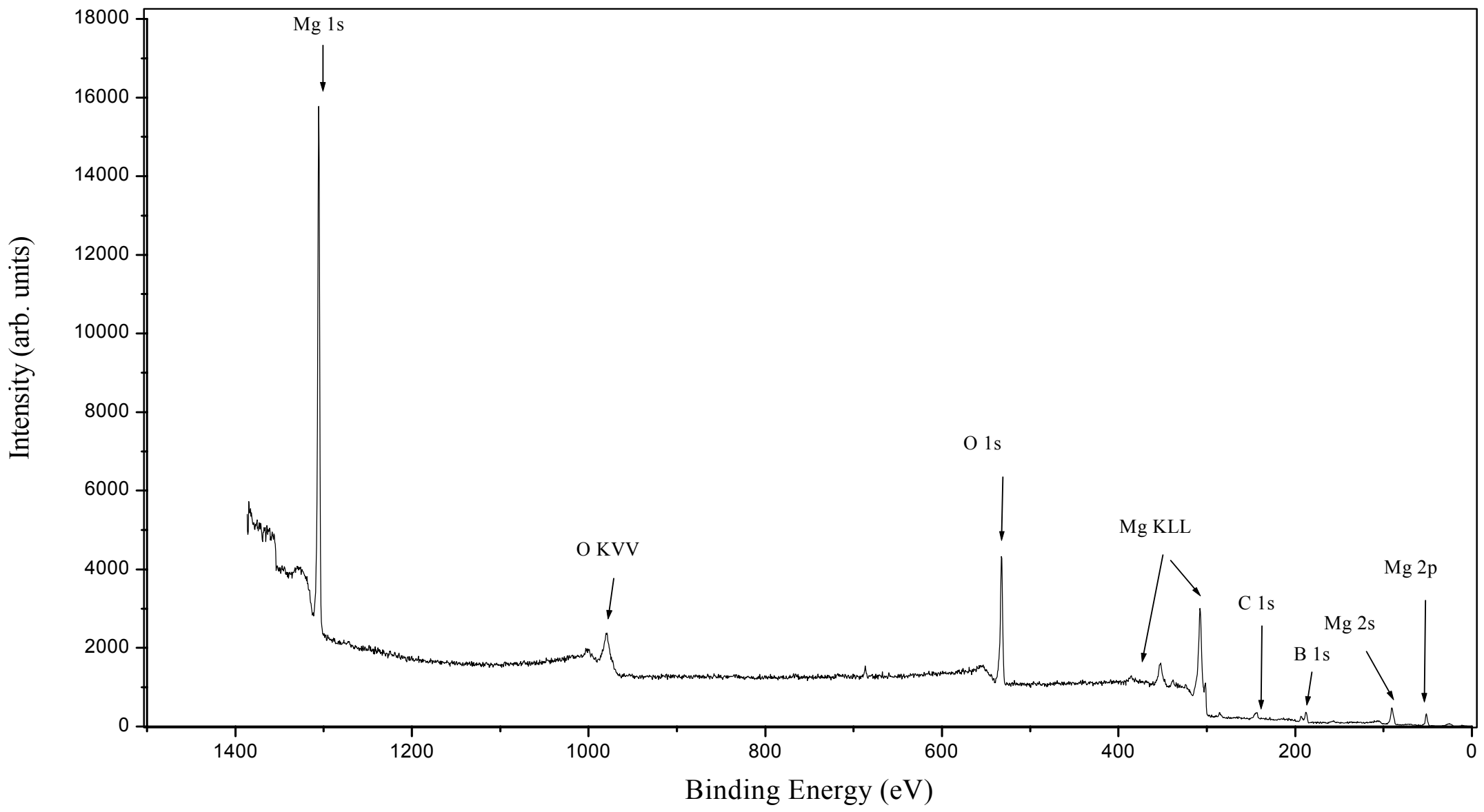

Fig. 3

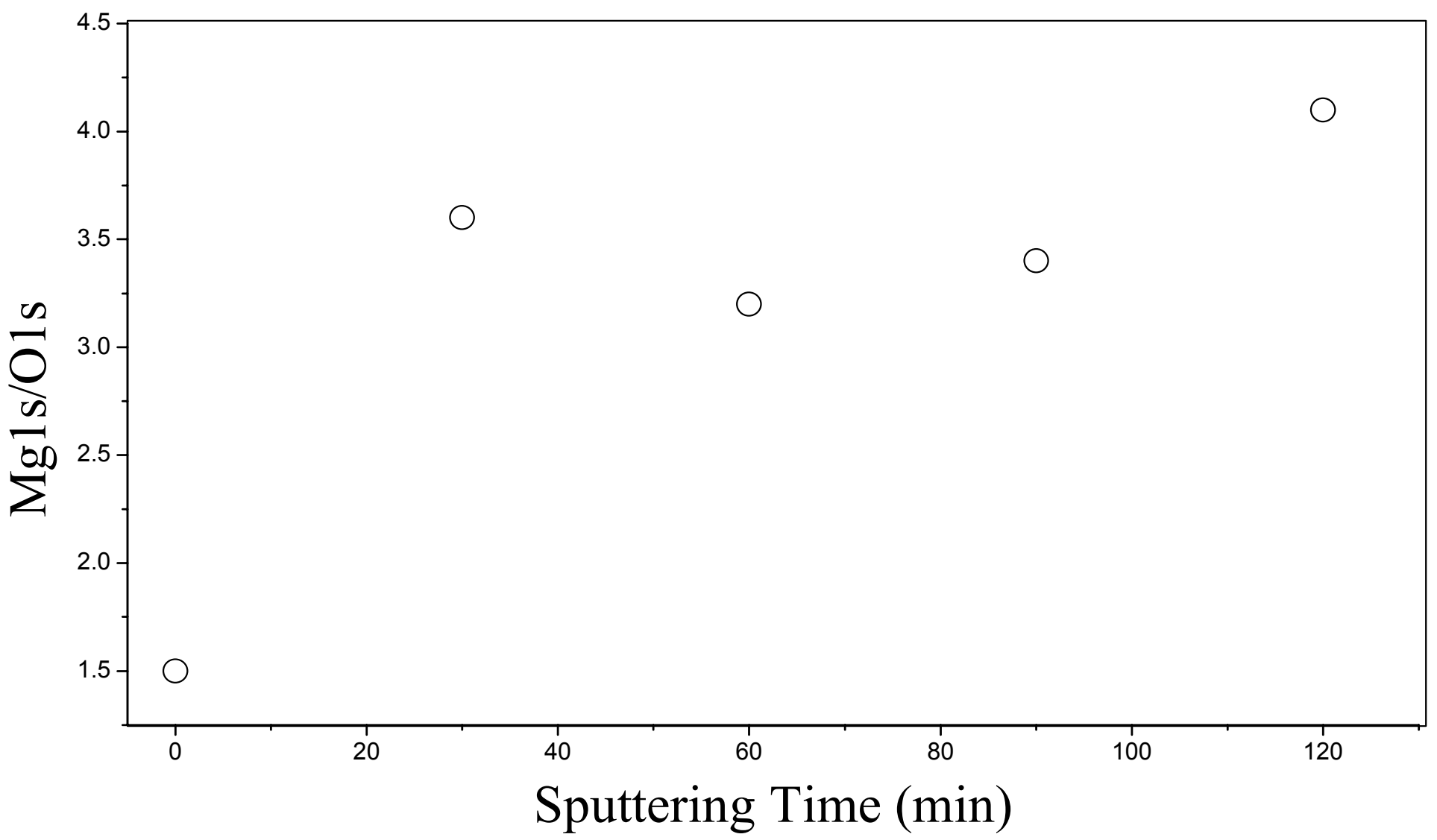

Fig 4

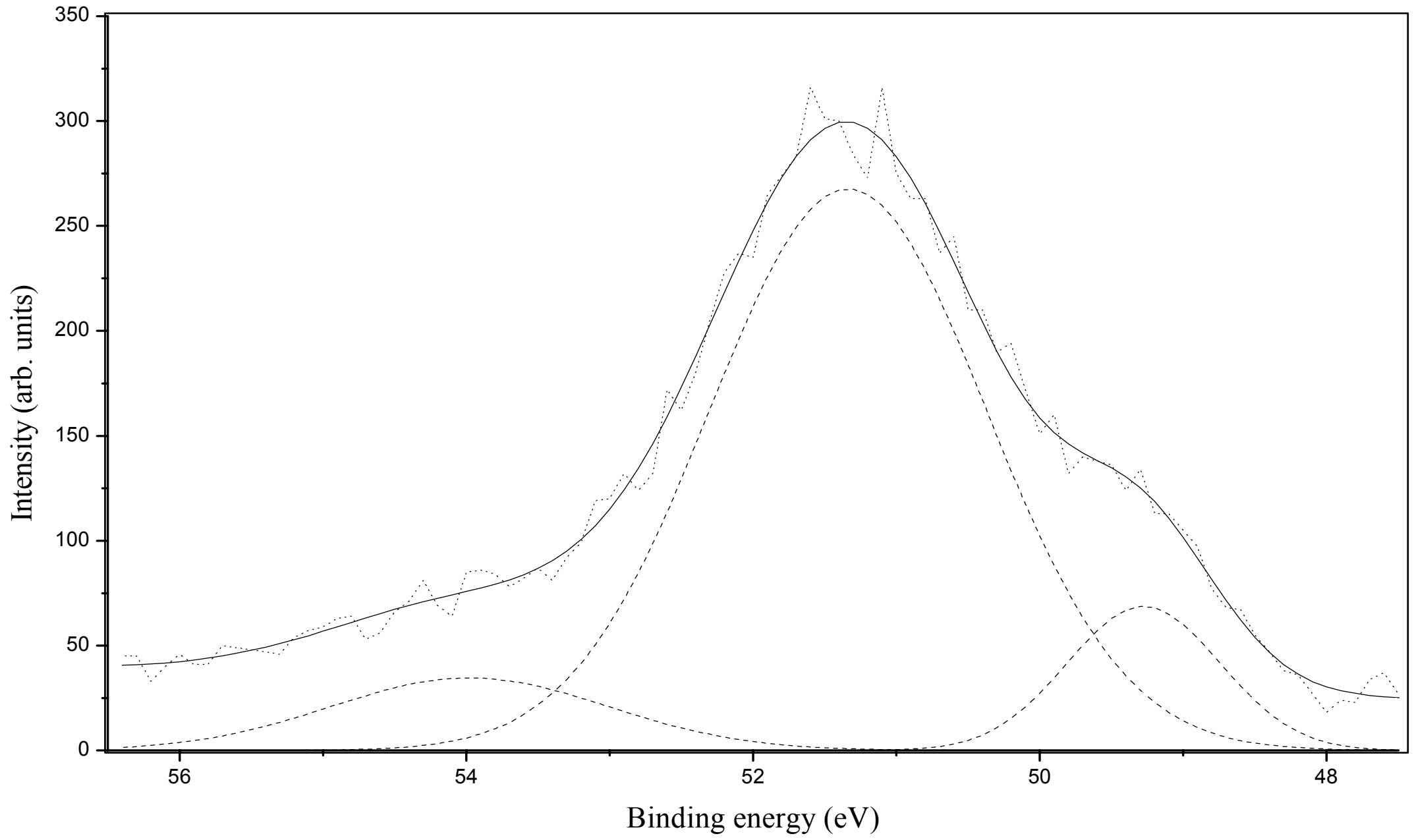

Fig. 5

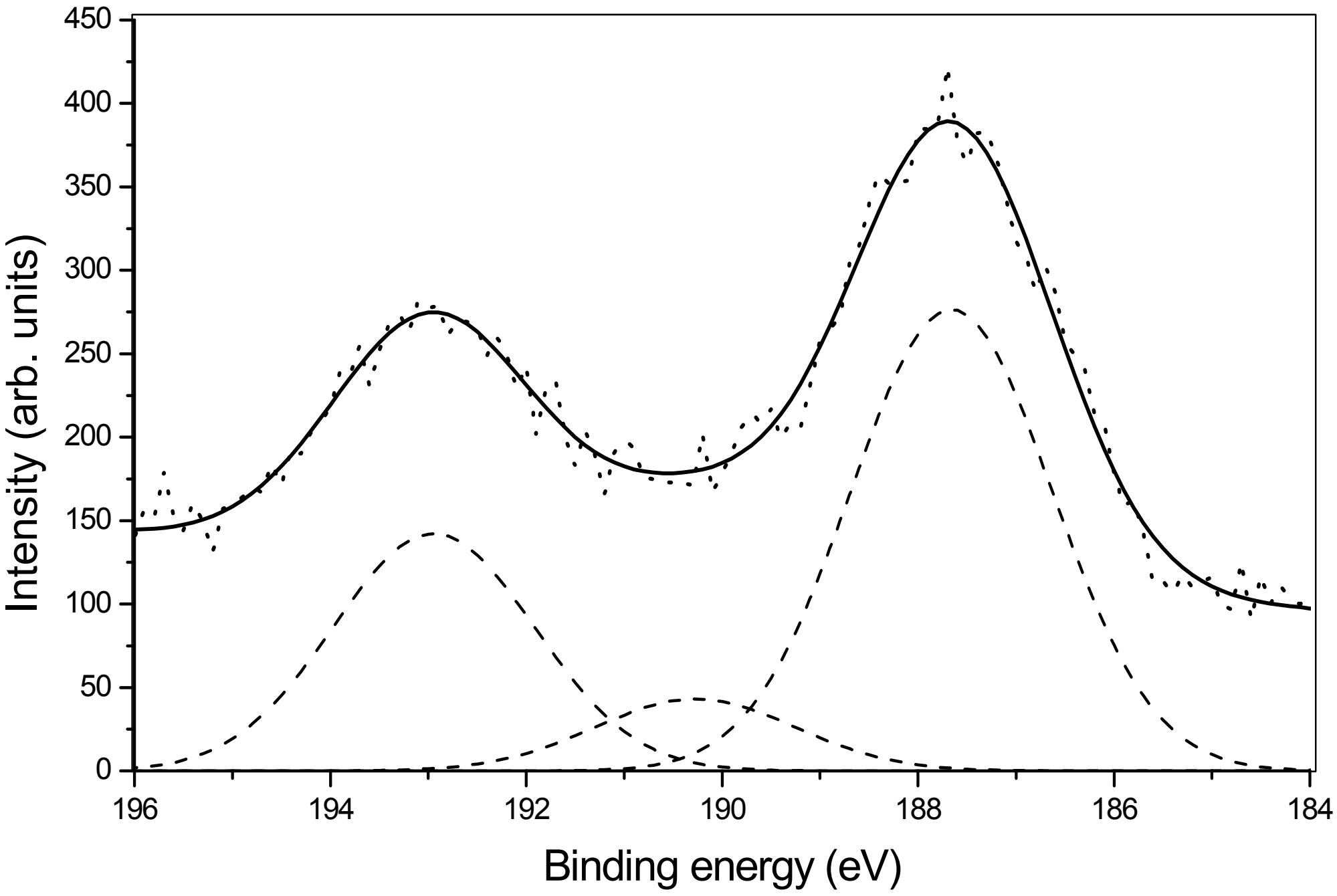

Fig. 6

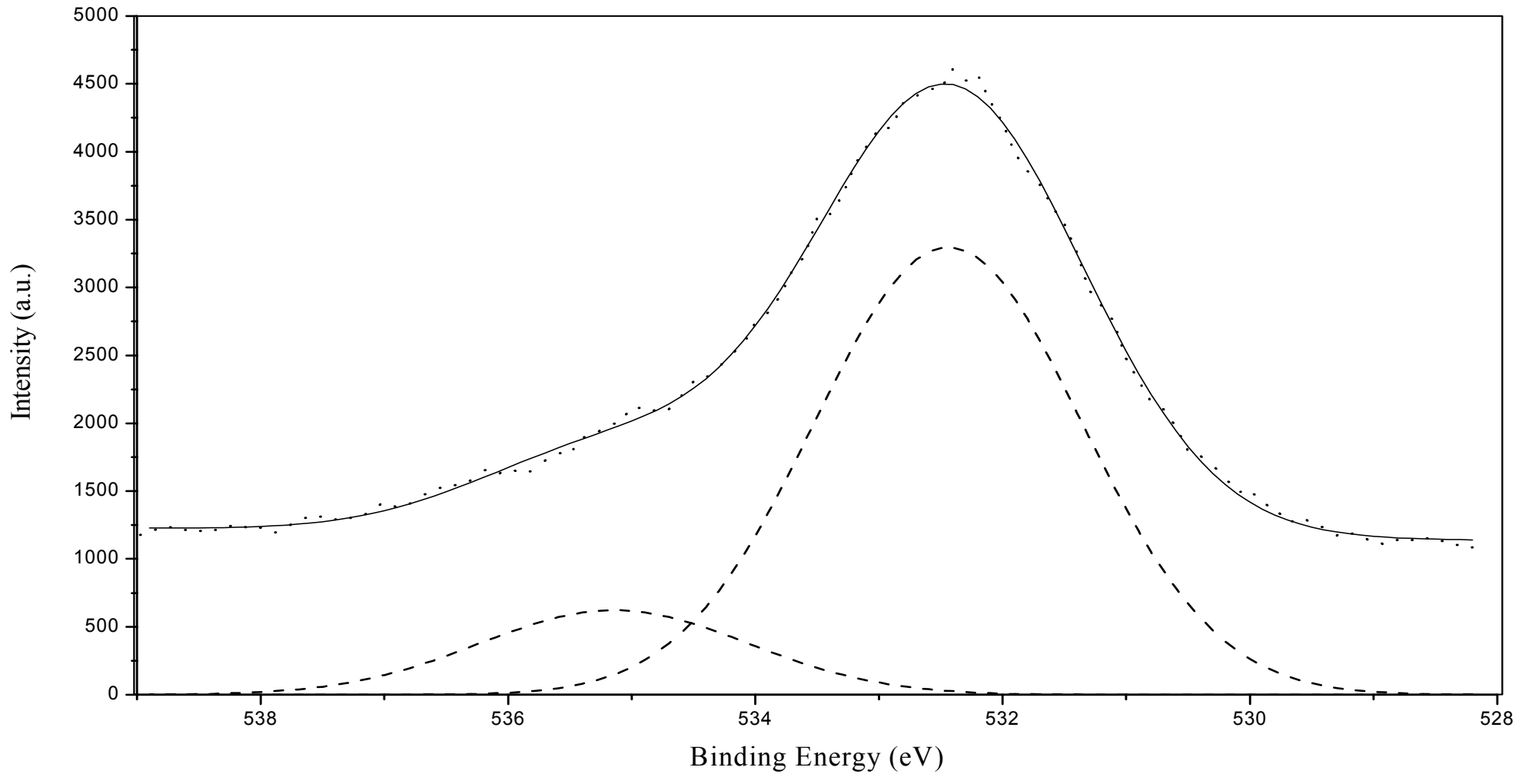

Fig. 7